\documentstyle[11pt,newpasp,twoside,epsf]{article}
\def\edcomment#1{\iffalse\marginpar{\raggedright\sl#1\/}\else\relax\fi}
\reversemarginpar

\newcommand{\approxlt}{\mbox{$\;^{<}\hspace{-0.24cm}_{\sim}\;$}}

\begin{document}
\title{A Search for Sub-millisecond Pulsations in Unidentified
FIRST and NVSS Radio Sources}
\author{Fronefield Crawford, Victoria M. Kaspi}
\affil{Center for Space Research, MIT, Cambridge, MA 02139, USA}
\author{Jon F. Bell}
\affil{ATNF, CSIRO, P.O. Box 76, Epping, NSW, Australia}

\begin{abstract}
We have searched 92 unidentified sources from the FIRST and NVSS radio
catalogs for pulsations at 610 MHz. The selected sources are bright,
have no identifications, are unresolved and have significant linear
polarization. Our search was sensitive to sub-millisecond pulsations
from pulsars with a large range of 
dispersion measures. We have detected no radio pulsations from
these sources and conclude that they are unlikely to be a population
of previously undetected pulsars.
\end{abstract}

\noindent The FIRST and NVSS surveys are recent 1400 MHz VLA radio
surveys of the Northern sky. In the published FIRST catalog (White et
al.~1997), the positions and fluxes of $\sim 1.4 \times
10^{5}$ discrete radio sources are complete to $\sim$ 1 mJy. The NVSS
survey (Condon et al.~1998) catalogs more than 1.8 $\times$ 10$^{6}$
sources complete to $\sim$ 2.5 mJy and preserves polarization
information.
Several large-scale pulsar surveys have been conducted at
high Galactic latitudes (see Camilo 1997 for a review) which were
sensitive to sub-millisecond pulsars with only a small range of
dispersion measures (DM $\approxlt$ 10 pc cm$^{-3}$).  A targeted
search of unidentified FIRST and NVSS sources for sub-millisecond
pulsars with a wide range of DMs is feasible using small frequency
channel widths and a fast sampling rate.

We have searched for radio pulsations in 92 bright, point-like
unidentified sources from the FIRST and NVSS surveys which are more
than 5\% linearly polarized at 1400 MHz. Although there is no abrupt
cutoff separating the pulsar population from the extragalactic radio
source population, a polarization threshold of 5\% excludes most
($\sim$ 90\%) of the identified non-pulsar population while retaining
the majority ($\sim$ 90\%) of the identified pulsar population in the
NVSS survey (Han \& Tian 1999).  Each source was observed at a center
frequency of 610 MHz in two orthogonal linear polarizations for 420
sec with the Lovell 76-meter radio telescope. A bandwidth of 1 MHz was
split into 32 channels with signals from each channel recorded as a
continuous 1-bit digitized time series sampled at 50 $\mu$s.  We did
not detect any pulsations from the target sources. 

We argue that our
non-detections imply that the sources are unlikely to be pulsars.
The selected sources were bright, with the weakest
source having a 1400 MHz flux density of 15 mJy. Assuming a typical
pulsar spectral index of $\alpha = 1.6$ (Lorimer et al.~1995), this
source would have a flux density  
58 mJy at 600 MHz. This is significantly greater than
our sensitivity limit for the expected range of pulse periods and DMs
(see Fig.~1). All of our sources, therefore, were bright enough in the
absence of scintillation to be detectable.

Dispersion smearing is not a factor preventing detection, since all
but two of these sources have high Galactic latitudes ($|b| >
5^{\circ}$) and are expected to have DM $<$ 100 pc cm$^{-3}$
regardless of distance. This is well within our range of sensitivity
to sub-millisecond pulsations (DM $\approxlt 500$ pc cm$^{-3}$,
Fig.~1).  Interstellar scattering, estimated from the Taylor \& Cordes
(1993) model of the Galactic electron distribution, is 
expected to be negligible.

Scintillation could be preventing the detection of some of the sources
if they are pulsars closer than $\sim$ 2 kpc. This distance range is
reasonable: if the sources were pulsars at 2 kpc, their 400 MHz
luminosities would all be at the upper end of the pulsar luminosity
distribution ($L_{400} > 450$ mJy kpc$^{2}$). However, it is unlikely
that all of our sources would be suppressed in this way, so
scintillation cannot account for the non-detection of most of our
sources. We conclude that the question of the nature of these sources
remains open.  We have submitted an article to the Astronomical
Journal which details the results presented here.

\begin{figure}
\epsfxsize=3.45in
\centerline{\epsfbox{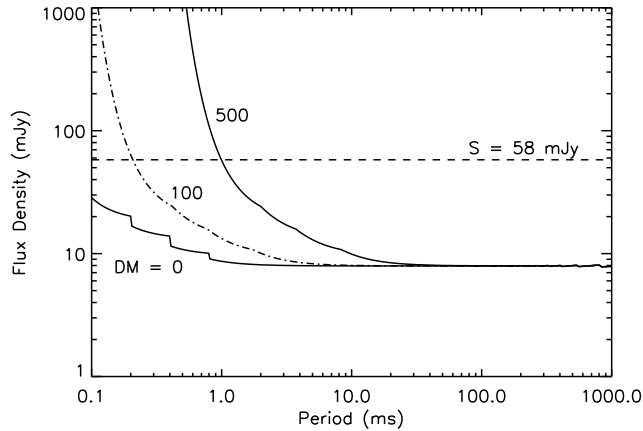}} 
\caption{Pulsar sensitivity curves for DMs of 0, 100, and 500 pc
cm$^{-3}$, assuming a pulsed duty cycle of 5\%.  The dashed horizontal
line at 58 mJy is the flux density of our weakest source at 600 MHz,
assuming a spectral index of $\alpha = 1.6$. All of our sources are
expected to have DM $<$ 100 pc cm$^{-3}$ (indicated by the
dashed-dotted line).}
\end{figure}

\references 

Camilo, F. 1997, in {\it High Sensitivity Radio Astronomy}, 
N. Jackson \& R. J. Davis, eds., 14 \\
Condon, J. J. et al. 1998, \aj, 115, 1693 \\
Han, J. L. \& Tian, W. W. 1999, \aap, 136, 571 \\
Lorimer, D., Yates, J., Lyne, A., \& Gould, D. 1995,
\mnras, 273, 411 \\
Taylor, J. H. \& Cordes, J. M. 1993, \apj, 411, 674 \\
White, R. L., Becker, R. H., Helfand, D. J., \& Gregg, M. D. 1997,
\apj, 475, 479

\end{document}